\documentclass[letterpaper]{article} 
\usepackage[submission]{aaai23}  
\usepackage{times}  
\usepackage{helvet}  
\usepackage{courier}  
\usepackage[hyphens]{url}  
\usepackage{graphicx} 
\usepackage{natbib}  
\usepackage{caption} 
\usepackage{algorithm}
\usepackage{algorithmic}
\usepackage{booktabs}
\usepackage[misc]{ifsym}

\usepackage{newfloat}
\usepackage{listings}
\DeclareCaptionStyle{ruled}{labelfont=normalfont,labelsep=colon,strut=off} 
\floatstyle{ruled}
\newfloat{listing}{tb}{lst}{}
\floatname{listing}{Listing}
\title{Multi-modal Protein Knowledge Graph Construction and Applications}
\author {
    Siyuan Cheng\textsuperscript{\rm 1,3},
    Xiaozhuan Liang \textsuperscript{\rm 1,3},
    Zhen Bi \textsuperscript{\rm 1,3},
    Huajun Chen \textsuperscript{\rm 2,3},
    Ningyu Zhang \textsuperscript{\rm 1,2,3} \footnote{Corresponding author.}
}
\affiliations {
    \textsuperscript{\rm 1} School of Software Technology, Zhejiang University, Hangzhou, China \\
    \textsuperscript{\rm 2} College of Computer Science and Technology, Zhejiang University, Hangzhou, China \\
    \textsuperscript{\rm 3} Alibaba-Zhejiang University Joint Research Institute of Frontier Technologies \\
    {\{sycheng, liangxiaozhuan, bizhen\_zju, zhangningyu, huajunsir\}}@zju.edu.cn
}

\usepackage{bibentry}

\begin{document}

\maketitle

\begin{abstract}
Existing data-centric methods for protein science generally cannot sufficiently capture and leverage biology knowledge, which may be crucial for many protein tasks. To facilitate research in this field, we create ProteinKG65, a knowledge graph for protein science. Using gene ontology and Uniprot knowledge base as a basis, we transform and integrate various kinds of knowledge with aligned descriptions and protein sequences, respectively, to GO terms and protein entities. ProteinKG65 is mainly dedicated to providing a specialized protein knowledge graph, bringing the knowledge of Gene Ontology to protein function and structure prediction. 
We also illustrate the potential applications of ProteinKG65 with a prototype. Our dataset can be downloaded at \url{https://w3id.org/proteinkg65}.
\end{abstract}

\section{Introduction}

Recent decades have witnessed the success of protein science with neural networks AlphaFold2 \cite{AlphaFold}, achieving remarkable performance in understanding the structure and functionality of the protein. However, relying solely on protein sequence information cannot adequately capture biological knowledge, which may be critical for many tasks. The effective use of knowledge in related fields can improve the upper limit of predicting protein structure and function. Note that biologists have contributed lots of domain knowledge in knowledge bases like Gene Ontology \cite{geneontology}.
Thus, it is intuitive to leverage knowledge graphs (KGs) as vital support for protein understanding.
Yet one major stumbling block is the limitation of high-quality knowledge graphs that can cover biology knowledge and protein sequences.

The research challenges of protein knowledge graph construction we face are as follows: various biological knowledge bases and protein data are publicly available in different formats on the Web; however, it is a non-trivial task to integrate those heterogeneous sources of data.
For example, Gene Ontology contains extensive biological functional knowledge but is not directly associated with protein sequences. And protein ID exists differently in additional protein databases. To integrate protein information, we propose \textbf{Uniprot \& GOA-based Protein Alignment}, a method that uses GOA and Uniprot ID mapping to align and merge data sources.
Another major challenge is the extremely unbalanced distribution. 
Statistically, we observe that 97.9\% of relational triples involve relations of {\itshape is\_a}, {\itshape part\_of} and {\itshape enables}, which makes it challenging to leverage long-tailed semantic knowledge for protein understanding. To alleviate this data imbalance, we propose \textbf{Gene Ontology-based Relation Refinement}, which uses the transitivity in the Gene Ontology annotation principles. 
For example, if a protein has an oxidoreductase function, it must also have a catalytic function. So we associate proteins with these high-level GO terms through their subfunctions.
We construct a multimodal protein KG based on the above-mentioned technologies and further discuss its potential applications.
The contributions of this study can be summarized as follows:
\begin{itemize}
    \item We contribute a multimodal protein knowledge graph containing all the experimentally verified high-quality proteins with biological knowledge.
    \item We provide a prototype with an online demo to predict the secondary structure and function based on ProteinKG65. For details, please refer to the demo website: \url{http://proteinkg.zjukg.cn}.
\end{itemize}

\begin{figure}[t]
    \centering
    \includegraphics[width=0.5\textwidth]{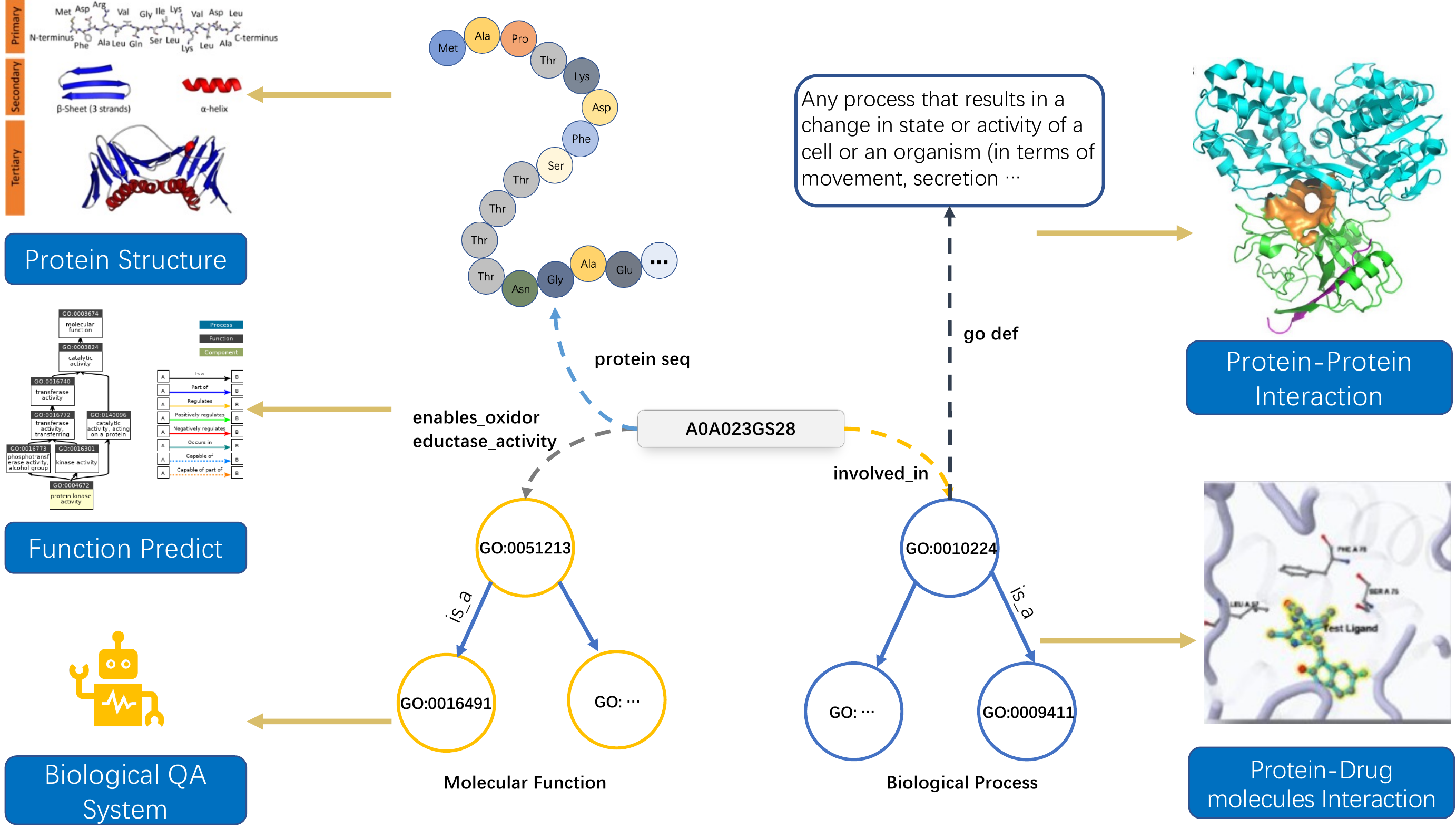}
    \caption{
    We show the format of ProteinKG65, taking the protein \texttt{A0A023GS28} as an example, showing some Go terms associated with this protein and their relationship and a wide range of application scenarios to ProteinKG65.}
    \label{main}
    \vspace{-1.0em}
\end{figure}

\section{ProteinKG65}
We build ProteinKG65 based on Gene Ontology\footnote{\url{http://geneontology.org}} and UniprotKB\footnote{\url{https://www.uniprot.org/}}. Figure \ref{main} shows the details of ProteinKG65 and some potential application scenarios. Gene Ontology describes the knowledge of the biological domain concerning molecular function, cellular components, and biological processes, which provides structured, computable knowledge about the function of genes and gene products. UniProtKB is the central hub for the collection of functional information on proteins with accurate, consistent and rich annotation.
\subsection{Multimodal Protein KG Construction}
\begin{figure}[t]
    \centering
    \includegraphics[width=0.5\textwidth]{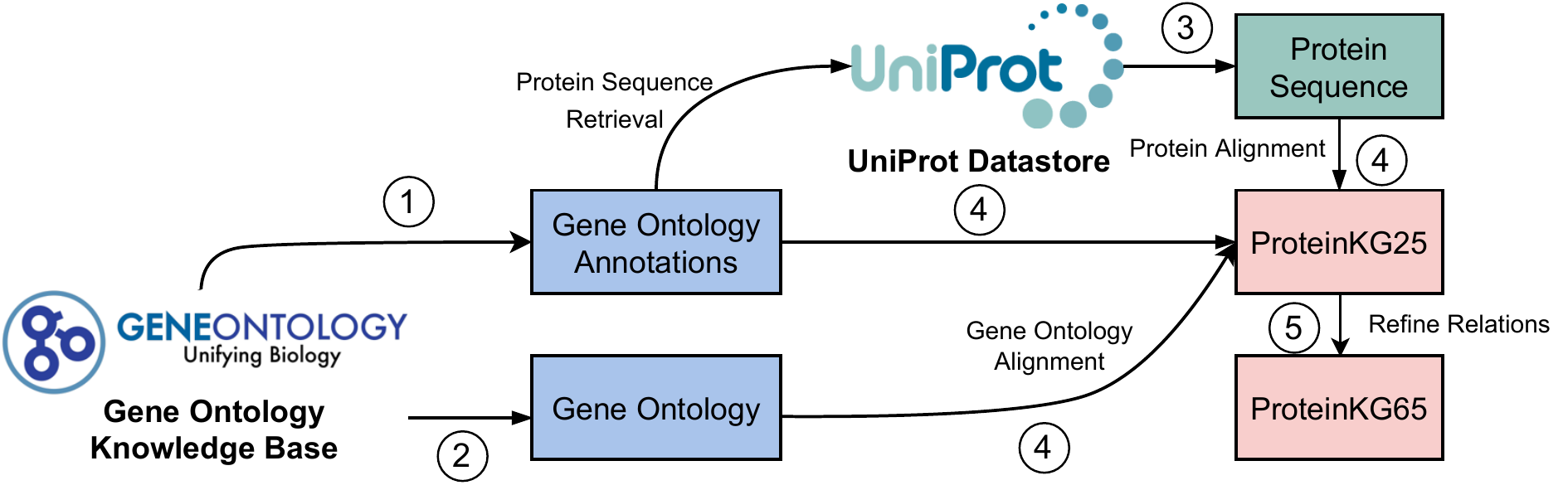}
    \caption{
    The construction process of ProteinKG65}
    \label{construct}
    \vspace{-1.0em}
\end{figure}
\quad \textbf{Uniprot \& GOA-based Protein Alignment:} In Figure \ref{construct}, we show the basic process of the method, the main steps are as follows:
Steps 1 and Step 2 extract gene annotations and gene ontology information from the gene ontology knowledge base. 
Step 3 and Step 4, we use the method to obtain all the proteins in the annotation and retrieve the corresponding protein sequence from the Swiss-Prot database. 
Then, we align and merge three data sources according to ID.

\textbf{Gene Ontology-based Relation Refinement:}Step 5, We refine the relations in ProteinKG25.
We sample GO terms from protein\--go triplets in three ontologies, denoted as:

\begin{equation}
     \mathcal{M} = \sum_{i}Sample(\mathcal{G}_{i}, k_{i}), i \in (BP, CC, MF)
\end{equation}
where $\mathcal{G}_{i}$ is GO terms in protein\--go triplets, and $k_{i}$ denotes the top $k_{i}$ nodes we want in different ontology.
After that, we try to extract the corresponding descendant GO sets of these terms, denoted as:
\begin{equation}
     \mathcal{C} = filter(\sum_{g} F(g)), g \in \mathcal{M}
\end{equation}
where function $F$ denotes getting the subtrees of nodes selected in $\mathcal{M}$. 
We use $filter()$ because there may be an overlap between the descendant sets of different GO terms in $\mathcal{M}$. 
$g$ is the terms name in sampled nodes. Finally, we update the original protein-go triplets, replacing the relation with a new generate relation $\mathcal{R}^N$ denoted as:

\begin{equation}
     \mathcal{R}^N =  \{ \mathcal{R}^O \oplus g | \mathcal{R}^{O}_{tail} \in \mathcal{C} , g \in \mathcal{M} \} \cup \{ \mathcal{R}^{O} | \mathcal{R}_{tail}^{O} \notin \mathcal{C} \}
\end{equation}
where $\mathcal{R}^O$ denotes the origin relation in the protein-go triplet. The GO appears in $\mathcal{C}$, we contact origin relation with top $k$ nodes selected in $\mathcal{M}$; otherwise, we keep the original.
\subsection{Statics \& Applications}
\begin{table}[t]
    \caption{
    The statistic of Protein-GO triplets in ProteinKG65. }
    \small
    \centering
    \renewcommand\tabcolsep{3.0pt}
    \begin{tabular}{lccccc}
    \toprule
    {Setting} & {} & {Protein} & {GO} & {Relation} & {Triplet} \\
    \midrule
   & train & 543,110 & 28,524 & 57 & 4,884,034 \\
   \specialrule{0em}{2pt}{2pt}
  Transductive & valid & 25,241 & 5,009 & 44 & 51,243  \\
  \specialrule{0em}{2pt}{2pt}
   & test & 217,463 & 17,908 & 57 & 575,160 \\
   \midrule
     & train & 543,110 & 28,524 & 57 & 4,884,034 \\
     \specialrule{0em}{2pt}{2pt}
  Inductive & valid & 855 & 270 & 31 & 2,216 \\
  \specialrule{0em}{2pt}{2pt}
   & test & 3,085 & 1,062 & 50 & 11,127 \\
    \bottomrule
    \end{tabular}
    \label{tab:Statistic}
    \vspace{-1.0em}
\end{table}

To make the ProteinKG65 consistent with the real-world application setting, we take two different settings for the protein-go triplets dataset(transductive \& inductive).
We detail the statistics of different settings in Table \ref{tab:Statistic}.

With the proposed ProteinKG65, we have implemented a prototype protein understanding system\footnote{\url{http://proteinkg.zjukg.cn}} based on a knowledge-enhanced pre-trained protein language model OntoProtein\footnote{\url{https://github.com/zjunlp/OntoProtein}}.
Biologists and computer scientists can utilize the prototype to analyze proteins for structure and function prediction.
Note that ProteinKG65 contains rich sequence–structure-function relations with biological expert experience, which can be applied to protein-drug molecule binding prediction and biological QA systems.

\section{Conclusion}
In this paper, we propose Uniprot \& GOA-based Protein Alignment and Ontology-based Relation Refinement to construct ProteinKG65 for protein science. 
We also provide the potential applications of ProteinKG65 with a prototype.
We hope our released knowledge graph can help to promote studies in AI for science.

\section*{Acknowledgment}

This work was supported by the National Natural Science Foundation of China (No.62206246), Zhejiang Provincial Natural Science Foundation of China (No. LGG22F030011), Ningbo Natural Science Foundation (2021J190), and Yongjiang Talent Introduction Programme (2021A-156-G), and CAAI-Huawei MindSpore Open Fund.

\bibliography{aaai23}
\end{document}